# Analysis of COVID19 Outbreak in India using SEIR model

Raj Kishore[1*], B. Sahoo[2], D. Swain[2], Kisor K. Sahu[1]

1. School of Minerals, Metallurgical and Materials Engineering, Indian Institute of Technology Bhubaneswar, India 752050
2. School of Earth Ocean and Climate Science Engineering, Indian Institute of Technology Bhubaneswar, India 752050

*Email: rk16@iitbbs.ac.in

**Abstract**

The prediction of spread patterns of COVID19 virus in India is very difficult due to its versatile demographic as well as meteorological data distribution. Various researchers across the globe have attempted to correlate the interdependency of these data with the spread pattern of COVID19 cases in India. But it is hard to predict the exact pattern, especially the peak in the number of active cases. In the present article we have tried to predict the number of active, recovered, death and total cases of COVID19 in India using generalized SEIR model. In our prediction, the occurrence of peak in the active cases curve has a very close match with the peak in the real data (difference of only one week). Although the number of predicted cases differs with the real number of cases (due to unlocking the movement restrictions gradually from June 2020 onwards), the close resemblance in the actual and predicted time (in the peak of active cases curve) makes this model relatively suitable for analysis of COVID19 outbreak in India.

## 1. Introduction

In December, 2019 in Wuhan city in Hubei province of the People's Republic of China (PRC) some pneumonia patients were reported [1-2]. But later it was found that the standard medical treatment protocol used for pneumonia on these patients was not effective and some of their conditions deteriorated rapidly. Thus it was declared that this is caused by a new virus, named as SARS-CoV-2 [3]. The reason behind such naming is due to the arrangement of the spike proteins of the virus that is indicative of a 'corona'. Since the virus responsible for the present epidemic related to the same family as that of SARS, so it was named as SARS-CoV-2 [4-5].

The initial outbreak started with the new year as per Chinese lunar calendar. Due to high human migration at that period due to festive season, the virus spreads quickly in China. Since humans are the major carrier of this virus, before it was noticed and get controlled, it silently dispersed across the entire globe. The 'success' of the virus is connected to its accidental capacity to exploit the human migration pattern. As we have already discussed that, at the early stage when the infection is largely limited to the upper respiratory track, the affected person mostly mistake its symptom as that of a mild flu and become contagious. In the absence of any clinical preventive mechanism (such as vaccine) or any effective drugs to cure the infected persons, containment of the disease through clinical interventions is still largely an unsolved puzzle. Therefore, the only possibility to contain the rapid spreading of the disease in communities is identifying and isolating such type of carriers by clinical diagnosis, which WHO referred to as "Test, test and test" [7]. Such type of strategy is effectively adopted by countries like South Korea and Singapore. However, for a large and highly populated country like India, there are operational, clinical, infrastructural and financial limitations towards adopting this kind of strategy at least at the early stage. So the other option is to deny an easy route to the virus that it can thrive on. Therefore, India took an unprecedented step of announcing a country wide 'complete lockdown' for 21 days, starting from 25th March, 2020 for its entire population of roughly 1.35 billion [8-9]. Meaning that, during this period the entire population were asked to remain confined within their home, or wherever they stayed at that point of time and all kinds of movements were largely prohibited except only for a tiny fraction responsible for providing essential services.

There are huge biological and medical research going on for finding the vaccine for this "unstoppable" epidemic [10-12]. But in this anti-epidemic battle, along with medical and biological research, theoretical research can also be very useful tool which uses statistical and mathematical modelling. It can be used for mapping the outbreak characteristic and forecasting the peaks and end time as well. For this purpose, several efforts have been made for calculating the several key parameters such as doubling time, reproduction rate, inflection point etc. [13-16]. The use of mathematical modelling based on dynamic equations [17-19] which uses time-series data is best suited for such scenario. One such widely used model is *Susceptible exposed infectious recovered model* termed as SEIR model [20-23]. The present article is based on one such theoretical study using generalized SEIR model which is the improvised version of classical SEIR model [20-21]. It includes two new states; the quarantined and insusceptible cases [24]. These includes the effect of preventive measures taken at early stages like confining into closed boundaries, wearing masks and maintaining social distancing etc. The brief description of the model is given in the following section. We have predicted the outbreak of COVID19 in India in between 10th June 2020 to 7th June 2021, using the real data available in between 15th April to 9th June 2020. The occurrence of peak in the predicted curve of total active cases is closely matching with the real curve of active cases with the difference of only one week.

## 2. Method

### 2.1 Model Description

The classical SEIR model was generalized and used for characterizing the COVID-19 outbreak in Wuhan, China at the end of 2019 by L. Peng *et al*. [24]. This model consist of seven different parameters namely $S(t), E(t), I(t), R(t), Q(t), D(t)$ and $P(t)$ which are varying with time $t$, and represent the respective numbers of *susceptible cases* (peoples which are having chances to get infected), *exposed cases* (peoples which are having virus in their body but still not capable of spreading the disease), *infected cases* (peoples which are capable of spreading the disease), *quarantined cases*(peoples which are infected but isolated), *recovered cases*, *death cases*, and *insusceptible cases*(peoples which are having zero chances of getting infected due to either isolated initially or following the rules like using regular face mask, social distancing, regular hand wash etc.) respectively. The relation between these seven parameters are shown in Fig. 1. These relations can be also represented mathematically in the form of ordinary differential equations (ODE) as shown in eq. (1-7).

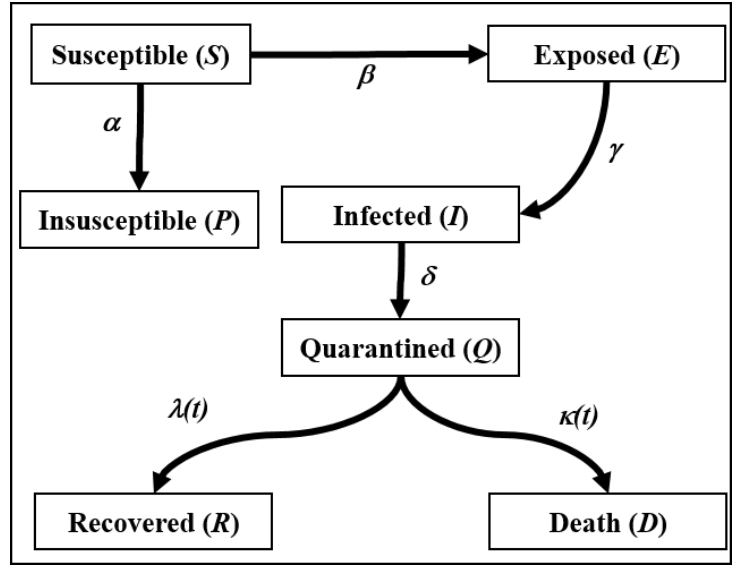

Fig. 1: The interconnectivity of different states of generalized SEIR model

The coefficients used in these ODE's α, β, γ, δ, λ(t), κ(t) are protection rate, infection rate, inverse of the average latent time, rate at which infectious people enter in quarantine, time-dependent recovery rate, time-dependent mortality rate respectively.

$$dS(t)/dt = -\beta\, S(t)I(t)/N - \alpha S(t) \tag{1}$$

$$dE(t)/dt = \beta\, S(t)I(t)/N - \gamma E(t) \tag{2}$$

$$dI(t)/dt = \gamma E(t) - \delta I(t) \tag{3}$$

$$dQ(t)/dt = \delta I(t) - \lambda(t)Q(t) - \kappa(t)Q(t) \tag{4}$$

$$dR(t)/dt = \lambda(t)Q(t) \tag{5}$$

$$dD(t)/dt = \kappa(t)Q(t) \tag{6}$$

$$dP(t)/dt = \alpha S(t) \tag{7}$$

The term $N$ represent the total population ($N = S+E+I+R+Q+D+P$) and assumed as constant which means that the births and natural deaths are not modelled here. It is to be noted that the recovery and mortality rate is time-dependent. It is due the behavior of recovery and death curve in the real data. From [25, 26], one can find that initially the recovery rate is low and gradually it increases over time whereas the mortality rate gradually decreases.

## 2.2 Parameter estimation

As the value of parameters α, β, γ, δ, λ(t), and κ(t) can greatly affect the final outcome of the model, the parameter estimation is very important step in such kind of theoretical study. Their values are estimated by fitting the available data. The best fitted parameters value in the present study is given in the Table 1. The mortality and recovery rate calculation is improvised by [27] and is modelled as

$$\kappa(t) = \frac{\kappa_0}{\exp(\kappa_1(t - \tau_\kappa))) + \exp(-\kappa_1(t - \tau_\kappa)))} \tag{8}$$

or as

$$\kappa(t) = \kappa_0 \exp\left(-[\kappa_1 (t - \tau_\kappa)]^2\right) \tag{9}$$

or as

$$\kappa(t) = \kappa_0 + \exp(-\kappa_1(t + \tau_\kappa)) \tag{10}$$

Where $k_0$, $k_1$ and $\tau_k$ are parameters to be empirically determined. The parameters $k_0$ and $k_1$ have the dimension of the inverse of a time and $\tau_k$ has the dimension of a time. The idea behind using the format of eq. (8-10) is to decrease the mortality rate over time, which is evident from the real data [25, 26]. The selection of best mortality rate among the given three is based on the best curve fitting criterion. The function which gives the minimum error between the actual and the predicted data points is considered as the best mortality rate function.

Similarly, the recovery rate $\lambda(t)$ is either modelled as

$$\lambda(t) = \frac{\lambda_0}{1 + \exp(-\lambda_1(t - \tau_\lambda)))} \tag{11}$$

or as

$$\lambda(t) = \lambda_0 + \exp(-\lambda_1(t + \tau_\lambda)) \tag{12}$$

where $\lambda_0$, $\lambda_1$ and $\tau_\lambda$ are parameters to be empirically determined. The parameters $\lambda_0$ and $\lambda_1$ have the dimension of the inverse of a time and $\tau_\lambda$ has the dimension of a time. The idea behind assuming the cure rate function format as of eq. (11-12) is to make cure rate initially low but gradually increasing and finally becomes constant, which is similar as the real data [25, 26]. The selection of best cure rate from the given two rates in eq. (11-12) is again based on the best curve fitting criterion as discussed in mortality rate selection.

The numerical solution of given seven ODE's follows the following steps:

a) First transform the ODE's in the form $dY/dt = A * Y + F$ where, $Y=[S, E, I, R, Q, D, P]^T$, $A$ and $F$ are two matrices as given below

$$A = \begin{bmatrix} -\alpha & 0 & 0 & 0 & 0 & 0 & 0 \\ 0 & -\gamma & 0 & 0 & 0 & 0 & 0 \\ 0 & \gamma & -\delta & 0 & 0 & 0 & 0 \\ 0 & 0 & \delta & -\kappa(t)-\lambda(t) & 0 & 0 & 0 \\ 0 & 0 & 0 & \lambda(t) & 0 & 0 & 0 \\ 0 & 0 & 0 & \kappa(t) & 0 & 0 & 0 \\ \alpha & 0 & 0 & 0 & 0 & 0 & 0 \end{bmatrix} \quad F = S(t)*I(t) \begin{bmatrix} -\beta/N_{pop} \\ \beta/N_{pop} \\ 0 \\ 0 \\ 0 \\ 0 \\ 0 \end{bmatrix}$$

b) The equation $dY/dt = A*Y + F$ is then solved using fourth order *Runge-Kutta* method [28] for finding the values of **Y** matrix for next time step.

Table 1: The estimated value of parameters used in SEIR model

| Parameter | Optimized value (Fitted on data between 15th April to 09th June 2020) |
|---|---|
| $\alpha$ | 0.0097 |
| $\beta$ | 0.1423 |
| $\gamma$ | 0.1499 |
| $\delta$ | 0.0431 |

We have collected the data of number of infected, recovered and death cases of each state of India for each day starting from 15th April 2020 till 9th June 2020 from [25]. The data is processed and the respective total quarantined (Q), recovered (R) and death (D) cases in the entire country is calculated. The matlab code for SEIR model is available at [29] which was further modified and used for the Indian data.

**Results and discussion**

We have first fitted the active, recovered, death and total cases curve using the available data in between 15th April to 9th June 2020 (56 days). During the fitting process, the optimized value of parameters ($\alpha$, $\beta$, $\gamma$ and $\delta$) is calculated as well as the best function representing the mortality and recovery rate is selected from the given functions (eq. (8-12)). The fitted curve along with the actual curve for the active, recovered, death and total cases is shown in region (i) of fig. 2. Once the optimized value of these parameters is calculated, the fitted model is used for predicting the values of active cases, recovered, death and total cases for the future time interval. We have predicted these value from 10th June 2020 to 7th December 2021 (until the total number of active cases reduces to less than 1000) which is shown in region (ii) of fig. 2. In fig. 2, we can see clearly that the active case in India rises until 10th September 2020 and then it starts declining. Thus the peak in active cases is predicted in the second week of September 2020 based on the data available until 9th June 2020. It is interesting to note that the actual active case curve shown in fig. 3 has peak very close to the predicted curve in fig. 2. The difference in both the peak is only one week.

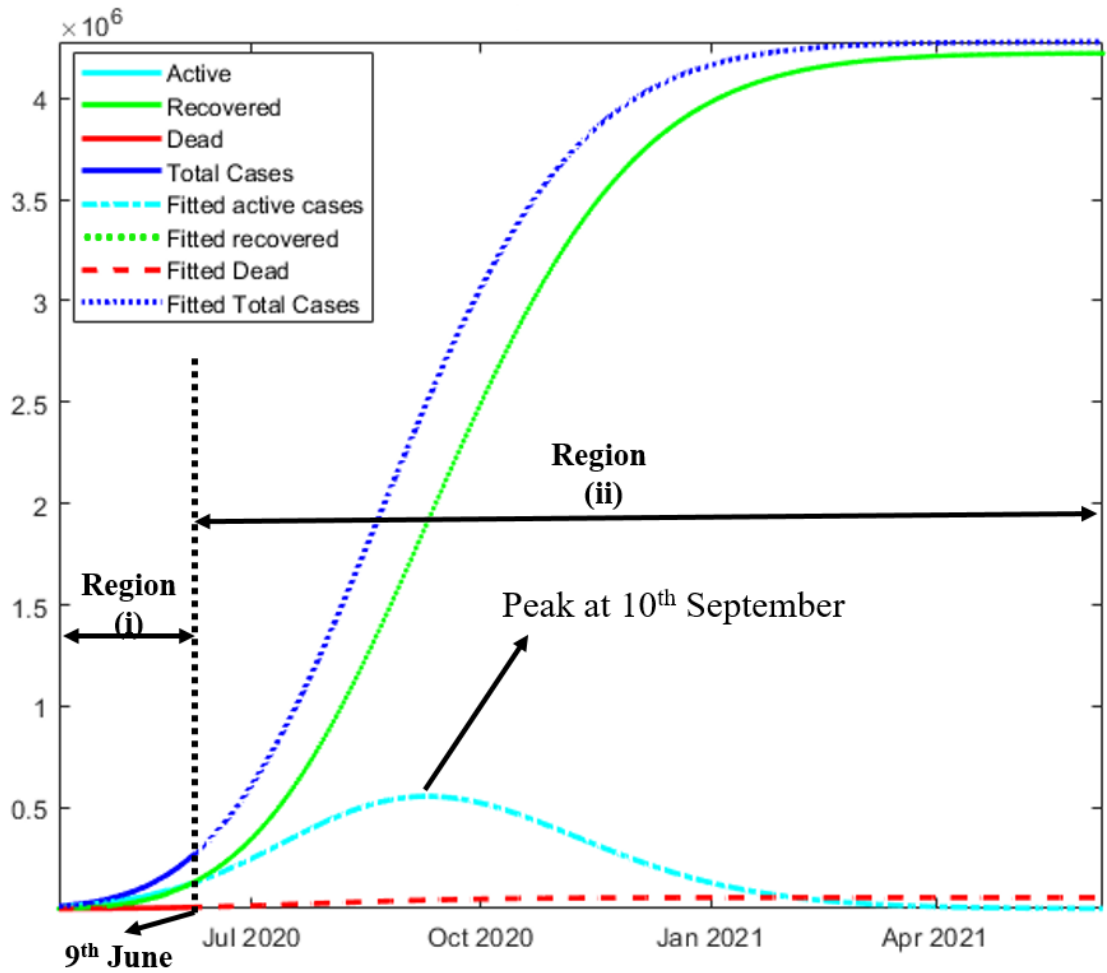

Fig. 2: The predicted values of active, recovered, death and total number of cases in between 10[th] June 2020 to 7[th] June 2021 (region (ii)). The peak in active cases occurs at 10[th] September 2020. The data used for fitting is in between 15[th] April to 9[th] June 2020 (region (i)).

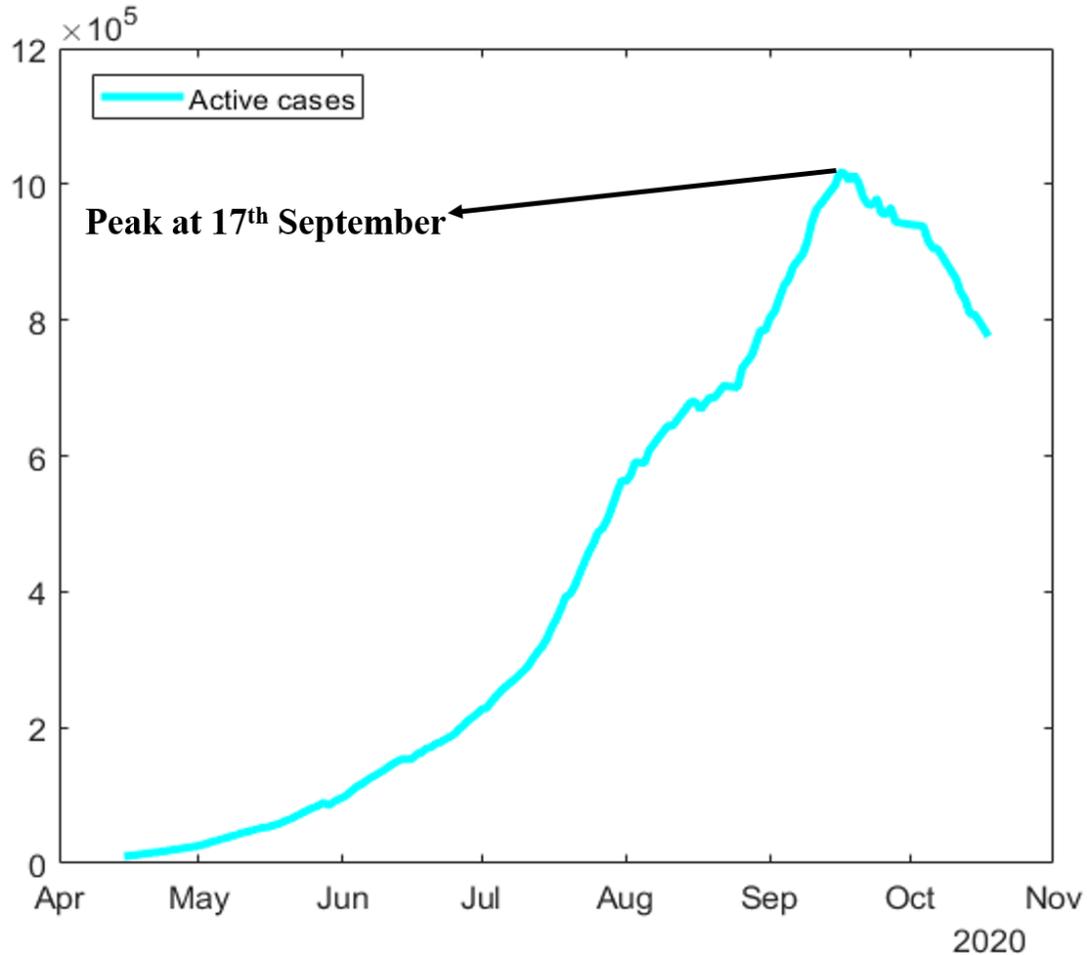

Fig. 3: The actual active cases of COVID-19 patients in between 15th April to 18th October 2020. The peak occurs at 17th September 2020. The data for active cases is taken from [25].

The number of recovered and total cases predicted using the model is lower than the actual values. The main reason behind this deviation is the unlocking the movement restrictions after 31st May 2020 [30]. Due to this decision of unlocking the country, taken by Indian government, the spread rate of the virus becomes way higher than that of during complete lockdown scenario. Due to this, the model which is fitted on the data from the complete lockdown period, predicted lower values for total number of new, recovered, and death cases.

**Conclusion**

The prediction of COVID19 outbreak in India is very difficult due to its vast demographic and meteorological data distribution. In the present article, we have tried to predict the peak and end time of the COVID19 cases in India using generalized SEIR model. The predicted time for the peak in the active cases is very close to the actual time of the peak in the active cases curve drawn using actual data. The difference in these two time interval is only one week. The model uses only data till 09th June 2020 and capable of predicting the peak which occurs in the month of September 2020. This suggest that the generalized SEIR model used in the present article is well suited for analyzing the COVID19 outbreak in India.